\documentclass[12pt]{article}
\usepackage[utf8]{inputenc}
\usepackage{amsmath}
\usepackage{amsfonts}
\usepackage{amssymb}
\usepackage{graphicx}
\usepackage{bm}% bold math
\usepackage{float}
\usepackage{ulem}
\usepackage{cancel} 
\usepackage{authblk}

\usepackage{slashed}
\usepackage{comment}
\usepackage{mathtools}
\usepackage[numbers,sort&compress]{natbib}

\newcommand{\mycomment}[1]{%
}%

\usepackage{graphicx}
\usepackage{fixltx2e,amsmath}
\usepackage[titletoc,toc]{appendix} 
\usepackage{extarrows}
\usepackage{bm}

\usepackage[colorlinks=true,
            linkcolor=OrangeRed,
            urlcolor=blue,
            citecolor=gray]{hyperref}
\hypersetup{citecolor=NavyBlue}  % MidnightBlue   NavyBlue

\usepackage[usenames,dvipsnames]{color}
\definecolor{dkgray}{RGB}{145,145,145}
\definecolor{violet}{RGB}{50,0,200}
\date{\today} %2022.XX.XX

\title{Random Magnetic Field and the Dirac Fermi Surface}
\author{Chao-Jung Lee}
\affil{\normalsize \it Department of Physics, California Institute of Technology, Pasadena, CA 91125, USA}
\author{Michael Mulligan}
\affil{ \normalsize \it Department of Physics and Astronomy, University of California, Riverside, CA 92511, USA}

\newcommand\dout{\bgroup \markoverwith{\rule[0.2ex]{0.1pt}{0.4pt}\rule[0.8ex]{0.1pt}{0.4pt}}\ULon}

\newcommand{\nn}{\nonumber}

\newcommand{\stkout}[1]{\ifmmode\text{\sout{\ensuremath{#1}}}\else\sout{#1}\fi}

\newcommand{\Tr}{\text{Tr}}

\newcommand{\com}[1]{}
\usepackage{leftidx}

\numberwithin{equation}{section}

\begin{document}

\maketitle
\thispagestyle{empty}

%	Cerulean blue	#2A52BE	\definecolor{ceruleanblue}{rgb}{0.16, 0.32, 0.75}
\definecolor{darkcerulean}{rgb}{0.03, 0.27, 0.49}
\definecolor{ceruleanblue}{rgb}{0.16, 0.32, 0.75}

%\title{Random Magnetic Field and the Dirac Fermi Surface}
%\author{Chao-Jung Lee}
%\affiliation{Department of Physics, California Institute of Technology, Pasadena, CA 91125, USA}
%\author{Michael Mulligan}
%\affiliation{Department of Physics and Astronomy, University of California, Riverside, CA 92511, USA}

\begin{abstract}
We study a single 2d Dirac fermion at finite density, subject to a quenched random magnetic field.
At low energies and sufficiently weak disorder, the theory maps onto an infinite collection of 1d chiral fermions (associated to each point on the Fermi surface) coupled by a random vector potential.
This low-energy theory exhibits an exactly solvable random fixed line, along which we directly compute various disorder-averaged observables without the need for the usual replica, supersymmetry, or Keldysh techniques. 
We find the longitudinal dc conductivity in the collisionless $\hbar \omega/k_B T \rightarrow \infty$ limit to be nonuniversal and to vary continuously along the fixed line. 
\end{abstract}

% {\color{BrickRed}Have to figure out if there is indeed $\frac{1}{N}$ for each $I,J...$ indices contraction ?  }
%%%%%%%%%%%%%%%%%%%%%%%%%%%%%%%%%%%%%%%%%%%%%%%%%%%%%%%%%%%%%%%%%%%%%%%%%%%%%%%%%%%%%%%%%
\definecolor{napiergreen}{rgb}{0.16, 0.5, 0.0}
\definecolor{officegreen}{rgb}{0.0, 0.5, 0.0}

\definecolor{seagreen}{rgb}{0.18, 0.55, 0.34}
\definecolor{sacramentostategreen}{rgb}{0.0, 0.34, 0.25}
\definecolor{upforestgreen}{rgb}{0.0, 0.27, 0.13}

\definecolor{tropicalrainforest}{rgb}{0.0, 0.46, 0.37}
%\definecolor{tropicalrainforest}{rgb}{0.0, 0.5, 0.3}

\definecolor{viridian}{rgb}{0.25, 0.51, 0.43}

\definecolor{pakistangreen}{rgb}{0.0, 0.35, 0.0}

\definecolor{GreenForTableofCont}{rgb}{0.0, 0.5, 0.32}

\newpage

\setcounter{page}{1}

 {  \hypersetup{linkcolor=darkcerulean}  %{darkcerulean =GreenForTableofCont
  \tableofcontents
 }

\newpage

\section{Introduction}

In this paper, we introduce and analyze a new effective theory for studying the influence of a quenched random magnetic field on a single Dirac fermion, moving in two spatial dimensions (2d) and placed at finite density.
Applications of this theory include:
graphene when restricted to a single valley \cite{CastroNeto2009} (where the vector potential disorder describes the effects of ripples in the graphene sheet \cite{PhysRevLett.97.016801, PhysRevLett.97.196804}) and the gapless surface states of a time-reversal invariant 3d topological insulator \cite{qhz2008};
an integer quantum Hall plateau transition for spinless electrons in a periodic potential \cite{PhysRevLett.61.2015, Ludwig_etal_1994}; and Dirac composite fermion mean-field descriptions of the half-filled Landau level and topological insulator surface state \cite{DTSon-2015,WangSenthilfirst2015, PhysRevB.95.205137, Seiberg:2016gmd}, other even-denominator metallic states \cite{PhysRevB.99.125135, PhysRevB.98.165137, PhysRevLett.122.257203}, and the superconductor-insulator transition \cite{Mike-PVL}.
In this last guise, the random vector potential arises from a random scalar potential perturbation to the dual electron system.

In contrast to a 2d nonrelativistic fermion \cite{BERGMANN19841, 0034-4885-56-12-001}, a single Dirac cone enjoys a sort of topological protection against localization \cite{PhysRevLett.99.146806, PhysRevLett.100.246806, PhysRevB.89.165136}: A free Dirac fermion cannot be gapped without breaking on average either time-reversal symmetry\footnote{Here we are explicitly referring to the symmetry of the Lagrangian for a single Dirac fermion.
This may be realized in a time-reversal invariant way as the surface state of time-reversal invariant topological insulator.
In a purely 2d theory, the simplest model breaks time-reversal microscopically \cite{NiemiRQ, Redlichparityshort, PhysRevLett.61.2015}; the resulting theory preserves a nonlocal particle-hole symmetry \cite{DTSon-2015,Seiberg:2016gmd}, which we are here viewing as an effective time-reversal symmetry.} or charge conservation; the random vector potential is unique (among random perturbations quadratic in the fermions) in that it respects a particle-hole symmetry \cite{Ludwig_etal_1994}.
The Dirac theory therefore gives rise to a delocalized, critical state for all values of the Fermi energy.
The problem here is to find an effective theory for this metal.

The conventional analytical approach to this problem, due to Pruisken and others \cite{Pruisken-prl, 1983JETPL..38..552K, Pruisken-RG2, WEIDENMULLER198787, DHLee,Lee-Kim} (see \cite{RevModPhys.80.1355, 2021AnPhy.43568439K} for reviews and \cite{PhysRevLett.98.256801, PhysRevB.100.235124, PhysRevB.104.125119} for specific studies of a disordered Dirac fermion), coarse grains away the elementary fermionic excitations in favor of a nonlinear sigma model with topological term, in which renormalization group fixed points are parameterized by the (disorder-averaged) Dirac fermion conductivity tensor.
Because the longitudinal resistivity $\sim 1/\sigma_{xx}$ plays the role of a coupling constant in this model, it is a challenge to quantitatively extend this description to the regime, $\sigma_{xx} \sim e^2/h$, relevant to experiment (e.g., \cite{PhysRevLett.102.216801, PhysRevB.81.033305}).
Tsvelik \cite{PhysRevB.75.184201} has conjectured a $PSL(2|2)_8$ Wess-Zumino-Novikov-Witten theory for the $\sigma_{xx} \sim e^2/h$ regime, based on its consistency with various numerical studies of the Chalker-Coddington model \cite{Chalker-Coddington, PhysRevB.54.8708} for the integer quantum Hall transition.
Unfortunately, there is as yet no consensus among the most recent numerical works (e.g., \cite{PhysRevB.99.121301, 2019PhRvB..99b4205Z, PhysRevLett.126.056802, PhysRevLett.126.076801, 2021AnPhy.43168560D}): For instance, there are statistically significant deviations among the various predictions for the localization length exponent (see Fig.~1 of \cite{2021AnPhy.43168560D}).
These deviations are understood to be the result of finite-size corrections to scaling \cite{2012IJMPS..11...60S, PhysRevLett.109.206804}; effects that are exacerbated by a leading irrelevant perturbation that is close to marginality. 
Motivated in part by the lack of numerical consensus, Zirnbauer \cite{ZIRNBAUER2019458, ZIRNBAUER2021168559} has recently proposed a fixed point description in terms of a CFT with only marginal perturbations.
Currently, this scenario appears to lack direct numerical support \cite{2021AnPhy.43168560D, 2021arXiv211209847D}.
An alternative study of the Dirac theory using (non-)Abelian bosonization by Ludwig et al.~\cite{Ludwig_etal_1994} makes use of an $SU(2)$ symmetry that is present only at zero density.
The zero density theory exhibits a fixed line (as a function of the disorder variance), along which the dynamical critical exponent varies continuously while the longitudinal conductivity is constant and of order $e^2/h$. 
Perturbation by a chemical potential within this treatment leads to difficulty:
The associated chemical potential deformation breaks the $SU(2)$ symmetry and causes a flow to strong coupling.\footnote{It may be surprising that the chemical potential perturbation---an operator quadratic in the fermions---leads to strong coupling. The bosonization used in \cite{Ludwig_etal_1994} to treat the 2d disordered theory maps the chemical potential operator to a nonlinear cosine term in the corresponding boson field with imaginary coefficient.}

Here we present a different approach to studying the finite density theory:
We first take the limit of the theory that focuses on the low-energy fluctuations about the Dirac Fermi surface before incorporating the disorder.
Taking inspiration from \cite{Ludwig_etal_1994}, the idea is to try to study the effects of disorder on the scale-invariant effective theory of (a system with) a Fermi surface \cite{Polchinski-EFT, Shankar-RMP}, rather than the ``microscopic" Dirac theory that includes both particle and antiparticle excitations.
We show how this approach allows for an exact solution of the effective theory---a random fixed line---provided the quenched vector potential is suitably long ranged and weak, and for direct calculations of the effects of disorder on various physical observables without the need for replicas, supersymmetry, or Keldysh formalisms.
In contrast to the zero density theory \cite{Ludwig_etal_1994}, we find the longitudinal  dc conductivity---calculated in the collisionless $\hbar \omega/k_B T \rightarrow \infty$ limit \cite{PhysRevB.57.7157}---to vary continuously along this finite-density fixed line.

The remainder of the paper is organized as follows.
In \S \ref{Free-Dirac-SetUp} we derive the effective theory of a Dirac Fermi surface in a random magnetic field. 
This theory takes the form of an infinite collection of 1d chiral fermions---one fermion for each point on the Fermi surface---coupled by the quenched vector potential disorder.
The large emergent symmetry $U(N)$ with $N \rightarrow \infty$ of the clean theory allows for an exact solution of the disordered theory, for each disorder realization, in which the randomness is removed entirely from the effective action by a gauge transformation.
Operators not invariant under the $U(N)$ symmetry, however, depend on the disorder.  
In \S \ref{disorderedobservablessection} we use the exact solution to directly compute various disorder-averaged physical observables.
We warm up by showing that the average fermion Green's function is short-ranged and that the system has a finite density of states.
We then turn to a calculation of the longitudinal conductivity, finding the conductivity to vary continuously with the strength of the disorder.
In \S \ref{discussion} we summarize and discuss possible directions of future work.
Appendix \ref{Appendix-DisAvg} contains relevant details for the calculations summarized in the main text.

\section{Low-Energy Effective Theory}    
\label{Free-Dirac-SetUp}

In this section, we derive the low-energy effective theory of a 2d Dirac fermion in a random magnetic field in terms of an infinite collection of 1d chiral fermions, coupled via a random vector potential. 

\subsection{Low-Energy Limit}
\label{lowenergytheory}

Our starting point is the theory of a two-component Dirac fermion $\Psi(t, {\bf x})$ at finite density, coupled to a quenched random $U(1)$ vector potential $A_j({\bf x})$:
\begin{align}
S = S_0 + S_1,
\end{align}
where $S_0$ is the action of a free Dirac fermion in 2d,\footnote{This form of the action corresponds to a relativistic one with $\gamma$ matrices: $\big(\gamma^0, \gamma^1, \gamma^2 \big) = \big( \sigma_3, i \sigma_2, - i \sigma_1 \big)$.} 
\begin{align}
\label{S0}
S_0 = \int dt d^2 {\bf x}\ \Psi^\dagger(t, {\bf x}) \big(i \sigma_a \partial_a + \mu \big) \Psi(t,{\bf x}),
\end{align}
and $S_1$ is the coupling of the Dirac fermion to the vector potential disorder,
\begin{align}
\label{S1}
S_1 = \int dt d^2 {\bf x} \ \Psi^\dagger(t, {\bf x}) \sigma_j A_j({\bf x}) \Psi(t, {\bf x}).
\end{align}
Above, ${\bf x} =  (x_1, x_2)$; $a \in \{0,1,2\}$; 
$\sigma_0$ is the $2 \times 2$ identity matrix and $\sigma_j$ for $j = 1, 2$ are standard Pauli matrices;
repeated indices are summed over unless otherwise specified.
The chemical potential $\mu$ is finite and nonzero, and we have set the velocity of the Dirac fermion to unity.
The vector potential $A_j({\bf x})$ is chosen to be a zero-mean Gaussian random variable.
We specify its disorder ensemble in \S \ref{disordersection}; for the present discussion, we require that $A_j({\bf x})$ is of sufficiently long wavelength that its Fourier transform $A_j({\bf q})$\footnote{Our Fourier transform convention: $A_j({\bf x}) = \int {d^2{\bf p}\over (2 \pi)} \, e^{i {\bf p} \cdot {\bf x}} A_j ({\bf p})$ and $\Psi(t, {\bf x}) = \int {d \omega d^2{\bf p}\over (2 \pi)^3} \, e^{- i \omega t + i {\bf p} \cdot {\bf x}} \Psi(\omega, {\bf p})$.} is only nonzero for wave vectors $|{\bf q}| \ll 2 |\mu |$.
This condition on the vector potential restricts to small-angle impurity scattering between fermionic excitations of nearby Fermi points.

Following \cite{Shamit-MirrorSym}, we derive the low-energy limit of this theory, in which we first focus on the low-energy excitations near the Fermi surface, defined by \eqref{S0}, and then incorporate scattering between Fermi points mediated by $A_j$ in \eqref{S1}. 
This order of limits assumes that the fluctuations of $A_j$ are sufficiently weak compared with $| \mu |$.

In Fourier space, the equation of motion following from \eqref{S0} is
\begin{eqnarray}
\label{eomdirac}
\big( (\omega+ \mu) \sigma_0 - \, p_j \sigma_j   \big) \, \Psi(\omega, {\bf p}) =0,
\end{eqnarray}
where $\Psi(\omega, {\bf p})$ is the Fourier transform of $\Psi(t, {\bf x})$ and ${\bf p} = (p_1, p_2)$.
The equation of motion \eqref{eomdirac} implies that the particle/antiparticle excitations have the dispersion relation, $\omega + \mu = \pm p$ with $p = | {\bf p} | \geq 0$.
As such, \eqref{eomdirac} can be rewritten as
\begin{eqnarray}
P^{(\mp)}({\bf p}) \Psi(\omega, {\bf p}) = 0,
\end{eqnarray}
where the projection matrices,
\begin{eqnarray}
P^{(\mp)}({\bf p})  \equiv \frac{1}{2} 
\Big( \sigma_0 \mp  \, \frac{ p_j \sigma_j }{ p }   \Big).
\end{eqnarray}
Using the projection matrices, we may write \eqref{S0} in Fourier space as
\begin{eqnarray}
\label{S0fourierfirst}
S_0 =
 \int {d\omega d^2{\bf p} \over (2 \pi)^3} \ 
 \Psi^\dagger(\omega, {\bf p}) \Big(
( \omega+ \mu -  p )
 P^{(+)}({\bf p})
 + ( \omega+ \mu + p )
P^{(-)}({\bf p}) \Big)  \Psi(\omega, {\bf p}).
\end{eqnarray}
For a spherical Fermi surface, we parameterize ${\bf p} = p \, \hat r_\theta \equiv p \, ( \cos \theta, \sin \theta )$ with $\theta \in [0, 2 \pi)$ labeling the points on the Fermi surface.
The projection matrices become
\begin{align}
P^{(\mp)}({\bf p}) = \frac{1}{2}
\begin{pmatrix}
1  & \mp  e^{-i \theta} \\
\mp e^{+ i \theta} & 1 
\end{pmatrix}
\end{align}
and we may expand $\Psi$ in terms of its particle ($R$) and antiparticle ($L$) excitations as
\begin{align}
 \label{Psi-Field}
\Psi(t, {\bf x})
& = \int {d\omega d^2 {\bf p} \over (2 \pi)^3} \, e^{- i \omega t + i {\bf p} \cdot {\bf x}} \Big( \frac{1}{\sqrt{2}} 
\begin{pmatrix}
e^{- i \theta}\\
 1
\end{pmatrix}
R_{\theta}(\omega, p) + \frac{1}{\sqrt{2}} 
\begin{pmatrix}
- e^{- i \theta}\\
 1
\end{pmatrix} L_{\theta} (\omega, p)
\Big),
\end{align}
where 
\begin{align}
P^{(+)}({\bf p}) \Psi(\omega, {\bf p}) & = \frac{1}{\sqrt{2}} 
\begin{pmatrix}
e^{- i \theta}\\
 1
\end{pmatrix} R_{\theta}(\omega, p), \\
\quad P^{(-)}({\bf p}) \Psi(\omega, {\bf p}) & = \frac{1}{\sqrt{2}} 
\begin{pmatrix}
- e^{- i \theta}\\
 1
\end{pmatrix} L_{\theta}(\omega, p). 
\end{align}
Inserting this expansion \eqref{Psi-Field} into \eqref{S0fourierfirst}, the free Dirac fermion action in Fourier space simplifies to 
\begin{align}
S_0 = \int {d\omega d^2{\bf p} \over (2 \pi)^3}
\Big( R^\ast_{\theta}(\omega, p) \big(\omega + \mu - p \big) R_{\theta}(\omega, p)
 + L^\ast_{\theta}(\omega, p) \big(\omega+ \mu + p \big) L_{\theta}(\omega, p) \Big).
\end{align}
For $\mu > 0$, particles ($R$) with momentum $p \sim k_F \equiv | \mu |$ are light and antiparticles ($L$) have energy $\omega \geq k_F$ ; for $\mu < 0$, the antiparticles are light and the particles are heavy.
Therefore, depending on the sign of $\mu$, the low-energy effective theory for excitations with $p \sim k_F$ only retains the particles or antiparticles.
Expanding about the Fermi momentum $k_F$,
\begin{align}
p = k_F + p_\perp, \quad | p_\perp | \ll k_F,
\end{align}
we have 
\begin{align}
d^2{\bf p} = p d p d \theta \approx k_F d p_{\perp} d \theta,
\end{align}
and, for $\mu > 0$, $S_0$ becomes
\begin{align}
\label{S0fouriersimple}
S_0 = \int {d\omega d p_\perp d \theta \over (2 \pi)^3} \ R^\ast_{\theta}(\omega, p_\perp) \big(\omega - p_\perp \big) R_{\theta}(\omega, p_\perp),
\end{align}
where we replaced $R_{\theta}(\omega, p) \rightarrow {1 \over \sqrt k_F} R_{\theta}(\omega, p_\perp)$.
For $\mu < 0$, we substitute $R \rightarrow L$ and $\omega - p_\perp \rightarrow \omega + p_\perp$.
Note that the deviation $p_\perp$ of the momentum about $k_F$ can be positive or negative.
Depending on the sign of $\mu$, we may interpret $S_0$ \eqref{S0fouriersimple} as an infinite collection of 1d chiral fermions $R_{\theta}(\omega, p_\perp)$.
We will only consider a single species of fermions with $\mu > 0$ so the low-energy theory will only involve $R$ fermions.

The vector field $A_j({\bf x})$ couples the chiral fermions $R_{\theta}(\omega, p_\perp)$ to one another according to $S_1$ \eqref{S1}.
To derive this coupling, we first decompose the Fourier transform of the vector potential $A_j({\bf q})$ in terms of its longitudinal $A_L({\bf q})$ and transverse $A_T({\bf q})$ components:
\begin{align}
\label{vectordecomposition}
A_j({\bf q}) = i {q_j \over q } A_L({\bf q}) + i \epsilon_{j k} {q_k \over q } A_T({\bf q}).
\end{align}
The momentum transfer,
\begin{align}
{\bf q} = {\bf p} - {\bf p}',
\end{align}
where ${\bf p} = (k_F + p_\perp) \hat r_{\theta}$ and ${\bf p}' = (k_F + p'_\perp) \hat r_{\theta'}$.
Since $p \sim p' \approx k_F$, we take 
\begin{align}
{q_j \over q} \approx { \hat r_{\theta} - \hat r_{\theta'} \over | \hat r_{\theta} - \hat r_{\theta'} | }.
\end{align}
Plugging the resulting decomposition \eqref{vectordecomposition} and the expansion \eqref{Psi-Field} into \eqref{S1}, we obtain the low-energy vector potential coupling for $\mu > 0$:
\begin{align}
\label{vectorcoupling}
S_1 = i \int {d \omega d^2 {\bf p} d^2 {\bf p}' \over (2 \pi)^5}\ R^\ast_{\theta}(\omega, p) R_{\theta'}(\omega, p') {\rm sign}(\theta - \theta') e^{{i \over 2} (\theta - \theta')} A_T({\bf p} - {\bf p}'),
\end{align}
where ${\rm sign}(X) = 1$ for $X \geq 0$ and ${\rm sign}(X) = -1$ for $X < 0$.
We have dropped a particle-antiparticle coupling between $R$ and $L$ fermions that is mediated by the longitudinal component $A_L$ of the vector potential, since such a term has an energy cost $\sim 2 k_F$.
The longitudinal component is therefore absent from this low-energy vector potential coupling.
For $\mu < 0$, \eqref{vectorcoupling} acquires an overall minus sign and we substitute $R \rightarrow L$.

In general, $A_T({\bf q})$ couples $R_\theta(\omega, p)$ and $R_{\theta'}(\omega, p')$ for arbitrary $\theta$ and $\theta'$.
By assuming that $A_T({\bf q})$ is only nonzero for $|{\bf q}| \ll 2 k_F$ (see \S \ref{disordersection} for further discussion), we may further simplify the effective coupling \eqref{vectorcoupling}.
For $|{\bf q}| \ll 2 k_F$, the momentum transfer ${\bf q} = {\bf p} - {\bf p}'$ can be approximated as
\begin{align}
\label{approxtransfer}
{\bf p} - {\bf p}' = \big(p_\perp - p'_\perp \big) \hat r_{\theta'} + k_F \big(\theta - \theta' \big) {d \over d \theta'} \hat r_{\theta'},
\end{align}
where $\big|p_\perp - p'_\perp \big| \ll 2 k_F$ and $\big|\theta - \theta' \big| \ll 2$.
Here, $\hat r_{\theta'} = ( \cos \theta', \sin \theta' )$ and ${d \over d \theta'} \hat r_{\theta'} = ( - \sin \theta', \cos \theta' )$ are orthogonal unit vectors about the Fermi surface point $\theta'$.
Defining
\begin{align}
\label{VandA}
V_{\theta \theta'}\big(p_\perp - p'_\perp \big) \equiv i\, k_F \, {\rm sign}(\theta - \theta')\, e^{{i \over 2} (\theta - \theta')}\, A_T({\bf p} - {\bf p}') ,
\end{align}
and given \eqref{approxtransfer}, the vector potential coupling \eqref{vectorcoupling} becomes
\begin{align}
\label{simplifiedlowES1}
S_1 = \int {d \omega d p_\perp d \theta d p'_\perp d \theta' \over (2 \pi)^5}\ R^\ast_{\theta}(\omega, p_\perp) R_{\theta'}(\omega, p'_\perp) V_{\theta \theta'}\big(p_\perp - p_\perp' \big).
\end{align}
(Recall the rescaling $R_\theta(\omega, p) \rightarrow {1 \over \sqrt k_F} R_\theta(\omega, p_\perp)$.)
Note that the same $A_T({\bf q})$ can enter different components of $V_{\theta \theta'}\big(p_\perp - p'_\perp \big)$ since a given ${\bf q}$ can correspond to the momentum transfer between distinct pairs of fermions.

We next perform the inverse Fourier transform\footnote{We have $R_\theta(t, z) = \int {d \omega d p_\perp \over (2 \pi)^2} \, e^{- i \omega t + i p_\perp z} R_\theta(\omega, p_\perp)$ and $V_{\theta \theta'}(z) = \int {d p_\perp \over 2 \pi} \, e^{i p_\perp z} V_{\theta \theta'} (p_\perp)$.} $(\omega, p_\perp) \rightarrow (t, z)$ on the fields appearing in the low-energy forms of $S_0$ \eqref{S0fouriersimple} and $S_1$ \eqref{simplifiedlowES1}.
For $\mu > 0$, $S = S_0 + S_1$ becomes
\begin{align}
\label{lowEbeforegauge}
S = \int {dt dz d\theta \over 2 \pi} \, R^\ast_{\theta}(t, z) i \big(\partial_t + \partial_z \big) R_{\theta}(t, z)  + \int {dt dz d \theta d \theta' \over (2 \pi)^2} \, R^\ast_{\theta}(t, z)  V_{\theta \theta'}(z) R_{\theta'}(t, z).
\end{align}
The first term describes an infinite collection of 1d chiral fermions, labeled by the Fermi point $\theta$.
The second term is a (quenched) random matrix interaction $V_{\theta \theta'}(z)$ between fermions associated to the Fermi points $\theta$ and $\theta'$.

\subsection{Electrical Current}

The electrical current (density) is 
\begin{align}
\label{currentdef}
J_j(t,{\bf x}) = \Psi^\dagger(t, {\bf x}) \sigma_j \Psi(t, {\bf x}).
\end{align}
We are interested in the contribution to this current that arises from the low-energy excitations near the Fermi surface when $\mu > 0$.
The simplest way to obtain this low-energy current is to replace in \eqref{S1} the quenched vector potential $A_j({\bf x})$ by a slowly varying background vector field ${\cal A}_j(t, {\bf x})$ and to take the low-energy limit described in the previous section.
(``Slowly varying" means $|q_0|, |{\bf q}| \ll k_F$.)
This produces the coupling (compare with \eqref{vectorcoupling}),
\begin{align}
\label{initialvectorcoupling}
S_2 = \int {d \omega d^2 {\bf p} d \omega' d^2 {\bf p}' \over (2 \pi)^6}\ R^\ast_{\theta}(\omega, p) R_{\theta'}(\omega', p') \Big( {e^{i \theta} + e^{- i \theta'} \over 2} {\cal A}_x(q_0, {\bf q}) + {e^{i \theta} - e^{- i \theta'} \over 2 i} {\cal A}_y(q_0, {\bf q})\Big),
\end{align}
where $q_0 = \omega - \omega'$ and ${\bf q} = {\bf p} - {\bf p}'$.
As before, $p = k_F + p_\perp$ and $p' = k_F + p'_\perp$.
The variational derivative $J_j(- q_0, - {\bf q}) = (2 \pi)^3 {\delta S_2 \over \delta {\cal A}_j(q_0, {\bf q})}$ gives the Fourier transform of the low-energy current:
\begin{align}
J_x(- q_0, - {\bf q}) & = \int {d \omega d^2 {\bf p} d^2 {\bf p}' \over (2 \pi)^3} \delta({\bf q} - {\bf p} + {\bf p}') R^\ast_{\theta}(\omega, p) \Big( {e^{i \theta} + e^{- i \theta'} \over 2} \Big) R_{\theta'}(\omega - q_0, p'), \\
J_y(- q_0, - {\bf q}) & = \int {d \omega d^2 {\bf p} d^2 {\bf p}' \over (2 \pi)^3} \delta({\bf q} - {\bf p} + {\bf p}') R^\ast_{\theta}(\omega, p) \Big( {e^{i \theta} - e^{- i \theta'} \over 2i} \Big) R_{\theta'}(\omega - q_0, p').
\end{align}

In computing the electrical conductivity, we will use a mixed Fourier space representation of the ${\bf q} = 0$ component of these currents.
Fourier transforming $J_j(- q_0, {\bf q} = 0)$ with respect to $q_0$,\footnote{We use our previous convention for the Fourier transform of $R_\theta(t, z)$ and $J_j(t, {\bf q}) = \int {d q_0 \over 2 \pi} e^{- i q_0 t} J_j(q_0, {\bf q})$.} we have
\begin{align}
\label{xcurrent}
J_x(t) \equiv J_x(t, {\bf q} = 0) = \int {dz d \theta \over 2 \pi} \, R_\theta^\ast(t,  z) \cos(\theta) R_\theta(t, z), \\
\label{ycurrent}
J_y(t) \equiv J_y(t, {\bf q} = 0) = \int {dz d \theta \over 2 \pi} \, R_\theta^\ast(t,  z) \sin(\theta) R_\theta(t, z).
\end{align}
$J_j(t)$ is the sum over the Fermi surface of the fermion density, weighted by $\cos(\theta)$ or $\sin(\theta)$, according to the current component $j$.

\subsection{Disorder Ensemble}
\label{disordersection}

We take the vector potential $A_j({\bf q})$ to be a zero-mean Gaussian random variable with variance:
\begin{align}
\label{vectordisorderdef}
\overline{A_j({\bf q}) A_k({\bf q}')} = g \delta_{i j} f(|{\bf q}|) \delta({\bf q} + {\bf q}').
\end{align}
Here, $g$ is a dimensionless constant that controls the overall scale of the fluctuations of the vector potential; $g\, f(|{\bf q}|)$ is a unit-normalized function with support $|{\bf q}| \ll k_F$.
For definiteness, we choose $f(|{\bf q}|) = \exp(- |{\bf q}|/M)$, where $M \ll k_F$ provides a smooth cutoff on the momentum transfer ${\bf q}$ in a scattering process.

We would like to determine what \eqref{vectordisorderdef} implies for the random matrix interaction $V_{\theta \theta'}(z)$.
We will argue that $V_{\theta \theta'}(z)$ gives rise to a local interaction in the effective 1d theory.
To this end, we consider the disorder average:
\begin{align}
\overline{V_{\theta_1 \theta_2}(z) V_{\theta_3 \theta_4}(z') } & = - k_F^2 {\rm sign}(\theta_1 - \theta_2) {\rm sign}(\theta_3 - \theta_4) e^{{i \over 2} (\theta_1 - \theta_2)} e^{{i \over 2} (\theta_3 - \theta_4)} \cr
& \times \int {d p_\perp d p'_\perp \over (2 \pi)^2} e^{i p_\perp z} e^{i p'_\perp z'} \overline{A_T({\bf q}) A_T({\bf q}')},
\end{align}
where (using our conventions from \eqref{approxtransfer})
\begin{align}
{\bf q} & = p_\perp \hat r_{\theta_2} + k_F (\theta_1 - \theta_2) {d \over d \theta_2} \hat r_{\theta_2} , \\
{\bf q}' & = p'_\perp \hat r_{\theta_4} + k_F (\theta_3 - \theta_4) {d \over d \theta_4} \hat r_{\theta_4}. 
\end{align}
We decompose $A_T({\bf q})$ in terms of its Cartesian components by replacing
\begin{align}
i \, {\rm sign}(\theta_1 - \theta_2) e^{{i \over 2} (\theta_1 - \theta_2)} A_T({\bf q}) & = {e^{i \theta_1} + e^{- i \theta_2} \over 2} A_x({\bf q}) + {e^{i \theta_1} - e^{- i \theta_2} \over 2 i} A_y({\bf q}) , \\
i \, {\rm sign}(\theta_3 - \theta_4) e^{{i \over 2} (\theta_3 - \theta_4)} A_T({\bf q}') & = {e^{i \theta_3} + e^{- i \theta_4} \over 2} A_x({\bf q}') + {e^{i \theta_3} - e^{- i \theta_4} \over 2 i} A_y({\bf q}') ,
\end{align}
and then use \eqref{vectordisorderdef} to find
\begin{align}
\label{exactVdisorderdef}
\overline{V_{\theta_1 \theta_2}(z) V_{\theta_3 \theta_4}(z') } & = {k_F^2 \over 2} \int {d p_\perp d p'_\perp \over (2 \pi)^2} e^{i p_\perp z} e^{i p'_\perp z'} f(|{\bf q}|) \delta({\bf q} + {\bf q}') \big( e^{i (\theta_1 - \theta_4)} + e^{- i (\theta_2 - \theta_3)} \big).
\end{align}
We will approximate \eqref{exactVdisorderdef} by 
\begin{align}
\label{approxtVdisorderdef}
\overline{V_{\theta_1 \theta_2}(z) V_{\theta_3 \theta_4}(z') } = g k_F \delta(\theta_1 - \theta_4) \delta(\theta_2 - \theta_3) f(z - z'),
\end{align}
where 
\begin{align}
\label{fdef}
f(z) = \int {d p_\perp \over 2 \pi} \, e^{i p_\perp z} f(p_\perp)
\end{align}
and it is to be understood that $|\theta_1 - \theta_2| \ll 1$ and $|\theta_3 - \theta_4| \ll 1$.
The disorder \eqref{approxtVdisorderdef} is local in $z$ since it depends on the relative coordinate $z-z'$.
We expect \eqref{approxtVdisorderdef} to be a good approximation to \eqref{exactVdisorderdef} at small momentum transfers and sufficiently low energies.
More precisely, we require $M/k_F \ll 1$ and that the cutoff on $p_\perp \in ( - \Lambda, \Lambda)$ satisfies $\Lambda/M \ll 1$. 

A detailed explanation for the approximation \eqref{approxtVdisorderdef} goes as follows. 
We begin by noting that for $M/k_F \ll 1$ and using $\delta({\bf q} + {\bf q}')$, $f({\bf q}) = f({\bf q}')$ fixes $|\theta_1 - \theta_2| \ll 1$ and $|\theta_3 - \theta_4| \ll 1$. 
Next, we expect the dominant contributions to \eqref{exactVdisorderdef} to arise when the terms in the sum $\big( e^{i (\theta_1 - \theta_4)} + e^{- i (\theta_2 - \theta_3)} \big)$ are in-phase.
This sets $\theta_1 = \theta_4 + \theta_3 - \theta_2 + 2 \pi m$, for an arbitrary integer $m$.
Using the earlier two conditions, the in-phase requirement implies $|\theta_1 - \theta_4 - \pi m| \ll 1$: This is modeled by $\delta(\theta_1 - \theta_4 - \pi |m|)$.
Substituting $\theta_1 = \theta_4 + \pi m$ into $f({\bf q})$ and $\theta_4 = \theta_1 - \pi m$ into $f({\bf q}')$ fixes $|\theta_4 - \theta_2 + \pi m| \ll 1$ and $|\theta_3 - \theta_1 + \pi m| \ll 1$.
Since $\theta_i \in [0, 2 \pi)$, these two conditions allow either $m = 0$ or $|m| = 1$.
First consider $m = 0$.
Since $\theta_2 \approx \theta_1$ and $\theta_4 \approx \theta_1$, ${\bf q} \approx p_\perp \hat r_{\theta_1} + k_F (\theta_1 - \theta_2) {d \over d \theta_1} \hat r_{\theta_1}$ and ${\bf q}' \approx p'_\perp \hat r_{\theta_1} + k_F (\theta_3 - \theta_4) {d \over d \theta_1} \hat r_{\theta_1}$.
Because $\hat r_{\theta_1}$ and ${d \over d \theta_1} \hat r_{\theta_1}$ are orthonormal, $\delta({\bf q} + {\bf q}') = \delta(p_\perp + p'_\perp) \delta\big( k_F (\theta_1 - \theta_2 + \theta_3 - \theta_4 ) \big) = {1 \over k_F} \delta(p_\perp + p'_\perp) \delta(\theta_3 - \theta_2)$, using $\delta(\theta_1 - \theta_4 - \pi |m|)$.
For Gaussian $f(|{\bf q}|)$, the dependence on $p_\perp$ and $\theta_1 - \theta_2$ factorizes.
Absorbing into $g$ the variation of this Gaussian on $|\theta_1 - \theta_2| \approx 0$, we perform the integral over $p'_\perp$ using $\delta(p_\perp + p'_\perp)$ to arrive at \eqref{approxtVdisorderdef}.
Next consider $m = 1$; the $m = -1$ case works similarly and will not be discussed.
Following the $m = 0$ logic, the replacement $\theta_4 = \theta_1 - \pi$ introduces a relative phase in the angular delta function,  $\delta({\bf q} + {\bf q}') = {1 \over k_F} \delta(p_\perp - p'_\perp) \delta\big(\theta_1 - \theta_2 - ( \theta_3 - \theta_4) \big) = {1 \over k_F} \delta(p_\perp - p'_\perp) \delta\big(2 \theta_1 - \theta_2  - \theta_3 - \pi\big) \approx {1 \over k_F} \delta(p_\perp - p'_\perp) \delta\big(\theta_1 - \theta_2 \big)$, using $\theta_3 \approx \theta_1 - \pi$.
The delta function has support when $\theta_1 = \theta_2$ (and similarly requires using the in-phase delta function $\delta(\theta_1 - \theta_4 - \pi)$ and $f({\bf q}')$ that $\theta_3 = \theta_4$).
For $\Lambda/M \ll 1$, scattering along the Fermi surface dominates and the relative contributions of $m \neq 0$ terms should be suppressed.
We therefore ignore the $|m| = 1$ terms in the remainder.

\subsection{Random Fixed Point and its Discrete Approximation} 
\label{exactsection}

Under the renormalization group transformation\footnote{The scale transformation is the following: $z \rightarrow \lambda z, t \rightarrow \lambda t, \theta \rightarrow \theta, R_\theta(t,z) \rightarrow \lambda^{-1/2} R_\theta(t,z)$.} \cite{Polchinski-EFT, Shankar-RMP} that leaves the $S_0$ part of \eqref{lowEbeforegauge} invariant, the leading flow equation \cite{PhysRevB.37.325} for the disorder variance $g$ is
\begin{align}
\label{deltargg}
{d g \over d \ell} = (3 - 2 \Delta) g,
\end{align}
where $\Delta = 1$ is the scaling dimension of $R_\theta^\ast(t,z) R_{\theta'}(t,z)$ and $\ell$ is the renormalization group length scale that increases as the energy is reduced.
We have substituted $f(z-z') = \delta(z - z')$ in deriving \eqref{deltargg}.
Randomness is therefore a relevant perturbation that drives the clean Dirac theory towards strong disorder.

Because the randomness is ${\cal O}(1)$ relevant, perturbation theory about the clean fixed point cannot access the strong disorder regime.
Luckily, the low-energy action \eqref{lowEbeforegauge} admits an exact solution for arbitrary $g$ (such that the derivation in \S \ref{lowenergytheory} holds), in which $V_{\theta \theta'}(z)$ is eliminated via the field redefinition \cite{Schaposnik-SchwingerModel, PhysRevB.51.13449, PhysRevD.53.3260}:
\begin{align}
\tilde R_{\theta}(t, z) = \int {d \theta' \over 2 \pi} U_{\theta \theta'}(z) R_{\theta'}(t, z),  
\end{align}
where the unitary matrix,
\begin{align}
\label{unitary}
U_{\theta \theta'}(z) \equiv \Big( {\cal T}_z e^{- i \int_{z_0}^z dz' V(z') } \Big)_{\theta \theta'}. 
\end{align}
Here, ${\cal T}_z$ denotes path ordering along $z$ and $z_0$ is an arbitrary base point.
The resulting action simplifies to 
\begin{align}
\label{gaugetransformedlowEaction}
S = \int {dt dz d\theta \over 2 \pi} \, \tilde R^\ast_{\theta}(t, z) i \big(\partial_t + \partial_z \big) \tilde R_{\theta}(t, z).
\end{align}
The action \eqref{gaugetransformedlowEaction} exactly describes the strong disorder regime of a Dirac fermion at finite density, subject to a random vector potential of sufficiently long wavelength.
The random vector potential has been eliminated from the effective action using the infinite-dimensional symmetry of the Fermi surface \cite{PhysRevX.11.021005, 2022arXiv220305004D, 2022arXiv220405328L}.

The formal manipulations above are made concrete by discretizing the $\theta$ coordinate (say, by putting the system in a finite-size box).
To this end, we take the Fermi surface to consist of $N$ discrete points: $\theta \rightarrow \theta_I = 2 \pi I/ N$ with $I = 1, \ldots, N$.
The fields and disorder are therefore replaced as
\begin{align}
R_\theta(t, z) & \rightarrow \sqrt N R_{\theta_I}(t,z) \equiv \sqrt N R_I(t,z) \\
V_{\theta \theta'}(z) & \rightarrow V_{\theta_I \theta_J}(z) \equiv V_{IJ}(z).
\end{align}
(The scaling of $R_{\theta_I}(t,z)$ by $\sqrt N$ is for notational simplicity.)
We are specifically interested in the limit $N \rightarrow \infty$.
Substituting in the discrete form of the angular integration $\int {d \theta \over 2\pi} \rightarrow {1 \over N} \sum_I$, we have the (equivalent) discrete forms for the low-energy action,
\begin{align}
\label{discreteone}
S & = \int dt dz \sum_{I=1}^N \, R^\ast_{I}(t, z) i \big(\partial_t + \partial_z \big) R_{I}(t, z)  + {1 \over N} \int dt dz \sum_{I, J = 1}^N \, R^\ast_{I}(t, z)  V_{I J}(z) R_{J}(t, z) \\
\label{discretetwo}
& = \int dt dz \sum_{I = 1}^N \, \tilde R^\ast_{I}(t, z) i \big(\partial_t + \partial_z \big) \tilde R_{I}(t, z),
\end{align}
where the rotated fermions, in discrete form, are
\begin{align}
\label{rotationdiscrete}
\tilde R_{I}(t, z) = \sum_{J = 1}^N U_{I J}(z) R_{J}(t, z), \quad U_{I J}(z) = \Big( {\cal T}_z e^{- {i \over N} \int_{z_0}^z dz' V(z') } \Big)_{IJ}.
\end{align}
The discrete action has an emergent $U(N)$ symmetry.
In discrete form, the low-energy currents \eqref{xcurrent} - \eqref{ycurrent} become 
\begin{align}
\label{discretexcurrent}
J_x(t) & = \int dz \sum_I R^\ast_{I}(t, z) \cos(\theta_I)R_{I}(t, z), \\
\label{discreteycurrent}
J_y(t) & = \int dz \sum_I R^\ast_{I}(t, z) \sin(\theta_I) R_{I}(t, z).
\end{align} 
For discrete $\theta$, the disorder variance \eqref{approxtVdisorderdef} becomes
\begin{align}
\label{approxtVdisorderdefdiscrete}
\overline{V_{I J}(z) V_{K L}(z') } = g k_F N \delta_{I L} \delta_{J K} f(z - z').
\end{align}
The overall factor of $N$ arises from the discrete form of $\delta({\bf q} + {\bf q}')$ in \eqref{exactVdisorderdef} with $\theta_I = 2 \pi I/N$.
It is the discrete form of the action \eqref{discretetwo} that we will use in the next section.

\section{Observables along the Fixed Line}
\label{disorderedobservablessection}

In general, in the presence of quenched disorder $V$ with unit-normalized distribution $P[V]$, the disorder-average of the correlation function of a physical observable ${\cal O}$ is defined as
\begin{align}
\label{Dis-Fun-Avg}
\overline{\langle {\cal O} \rangle}
\equiv
\int DV P[V] \, \langle  \mathcal{ O } \,  \rangle_{V}  ,
\end{align}
where the correlation function $\langle  \mathcal{ O } \,  \rangle_{V}$ in the disorder realization $V$ is
\begin{align}
\langle  \mathcal{O} \,  \rangle_{V} 
\equiv
\frac{  \int D\Phi\; \mathcal{O}   \; e^{i S[\Phi,V]}     }
{  \int D\Phi\; e^{ i S[\Phi,V]  }   }.
\end{align}
Here, we are momentarily denoting the dynamical fields of the theory by $\Phi$ and the action $S[\Phi, V]$ indicates the dependence upon both $\Phi$ and the disorder $V$.
In most theories, the presence of $V$ in the denominator of $\langle  \mathcal{O} \,  \rangle_{V}$ renders the direct analytic integration over all possible disorders in $\overline{\langle \mathcal{ O } \rangle}$ difficult, if not impossible.
As such, various ingenious tricks---such as replica, supersymmetric, and Keldysh formalisms---have been employed with various levels of success.
In this paper, we instead make use of the exact solution of the low-energy effective theory presented in \S \ref{exactsection} to directly perform the disorder average.

To see how this works, consider the correlation function of a local observable ${\cal O}(R)$, which is a function of the unrotated $R_I(t,z)$ fermion: 
\begin{eqnarray}
\label{Det-V-source-free}
\langle \mathcal{O}(R) \rangle_V
=
\frac{ \int D R^\dagger D R \; 
\mathcal{O}(R) \; e^{i S[R, V] }   }{ \int D R^\dagger D R \; e^{i S[R, V] }  }
=
\frac{ \int D \tilde R^\dagger D \tilde R \; 
\mathcal{O}(U^\dagger \tilde R) \; e^{i S[\tilde R] }   }{ \int D \tilde R^\dagger D \tilde R \; e^{i S[\tilde R] }  }.
\end{eqnarray}
In the first equality, $S[R,V]$ denotes the action \eqref{discreteone}; in the second equality, $S[\tilde R]$ denotes \eqref{discretetwo} with $\tilde R$ the rotated fermion \eqref{rotationdiscrete} with rotation $U$ a function of $V$.
After the rotation, the denominator no longer depends on $V$ and, in this case, yields the usual fermion determinant.
The disorder $V$ now only appears in the expression for the observable $\mathcal{O}(U^\dagger \tilde R)$ and may, in principle, be averaged over.\footnote{Note that any possible quantum anomalies \cite{Schaposnik-SchwingerModel} associated with the unitary rotations \eqref{unitary} of the chiral fermion path integral measures, being a function of the disorder $V$ only, mutually cancel between the numerator and denominator.}
The key point is that the local observables we consider separate into a sum of terms of the form,
\begin{align}
\label{factorization}
\mathcal{O}(U^\dagger \tilde R) =  \sum_{a,b} c_{ab} {\cal A}_a(\tilde R) \mathcal{B}_b(V),
\end{align}
for some constants $c_{ab}$, such that the disorder-averaged correlation function $\overline{\langle {\cal O}(R) \rangle}$ factorizes:
\begin{align}
\overline{\langle {\cal O}(R) \rangle} & = \sum_{a,b} c_{ab} \frac{ \int D \tilde R^\dagger D \tilde R \; 
\mathcal{A}_a(\tilde R) \; e^{i S[\tilde R] }   }{ \int D \tilde R^\dagger D \tilde R \; e^{i S[\tilde R] }  }  \cdot \int dV P[V] \mathcal{B}_b(V)\cr
& \equiv \sum_{a,b} c_{ab} \langle {\cal A}_a(\tilde R) \rangle \cdot \overline{ {\cal B}_b(V) }.
\end{align}
We will show how this can be used to compute the fermion Green's function and the longitudinal conductivity at the random fixed point \eqref{discretetwo}.

\subsection{Diffusive Green's Function and Density of States}

We begin by calculating the disorder-averaged Green's function and density of states.
We will find that the average Green's function is short-ranged and that the density of states is a positive constant.
These calculations will introduce the technique we will later use to calculate the conductivity.

Using the low-energy action \eqref{discretetwo}, the fermion two-point function averaged over the disorder is
\begin{align}
 \label{Greens-fun}
\overline{\langle  R_I(t,z) R^\dagger_I(0,0) \rangle}
& = \sum_{A,B}  \langle  \tilde R_A(t,z) \tilde R^\dagger_B(0,0) \rangle \cdot \overline{  U^\dagger_{IA}(z)  U_{BI}(0) } \cr
& = \sum_{A,B}  \frac{i}{2\pi}  \frac{ \delta_{AB}   }{(z - t) + i\alpha } \cdot \overline{  U^\dagger_{IA}(z)  U_{BI}(0) } ,
\end{align}
where $U(z)$ is defined in \eqref{rotationdiscrete}, $\alpha$ is a short-distance cutoff, and the index $I$ is not summed over.
Following \cite{Akuzawa-Wadati}, we compute the disorder average of $U$ matrices by decomposing the interval $[0,z]$ into $n$ steps $z_0 = 0, z_1 = \delta z, \ldots, z_n = n \delta z$ of length $\delta z = |z|/n$ such that
\begin{align}
\label{Udecomposition}
U(z) = e^{- {i \over N} \int_{z_{n-1}}^{z_n} dz' V(z') } \cdots e^{- {i \over N} \int_{z_{1}}^{z_2} dz' V(z') } e^{- {i \over N} \int_{z_{0}}^{z_1} dz' V(z') }.
\end{align}
We have chosen the arbitrary reference point $z_0 = 0$.
For sufficiently large $n$ with fixed $n \delta z = |z|$, the argument of each exponential can be approximated as
\begin{align}
{1 \over N} \int_{z_{j-1}}^{z_j} dz' V(z') \approx {1 \over N} V(z_j) \delta z \equiv M_j.
\end{align}
Using \eqref{approxtVdisorderdefdiscrete}, $M_j$ is a zero-mean Gaussian random variable with variance,
\begin{align}
\overline{(M_i)_{IJ} (M_j)_{KL}} = {g k_F \over N} \delta_{IL} \delta_{JK} f(z_i - z_j) \delta z^2,
\end{align}
where $f(z_i - z_j)$ is given in \eqref{fdef}.
For discrete $z$, we take
\begin{align}
\label{discretefdef}
f(z_i - z_j) \delta z = f_0 \delta_{|i - j|, 0} + f_1 \delta_{|i - j|, 1}.
\end{align}
The dimensionless coefficients $f_0$ and $f_1$ approximate a Gaussian $f(z_i - z_j)$ of finite width.

Using this, the disorder average of the product of $U$ matrices in \eqref{Greens-fun} becomes
\begin{eqnarray}   
\label{averageoftwoUs}
\sum_A  \overline{ U^\dagger_{IA}(z)  U_{AI}(0) }
  && = \sum_A
  \overline{   
 \Big( e^{ i \, M_1  } \cdots  e^{ i \, M_n }   \Big)_{IA}
 \times 
 \bm{1}_{AI} } \cr
&& =
     \overline{   
 \Big(  [ 1+ i M_1  -\frac{M_1^2}{2}+ \ldots]
  \cdots
 [ 1+ i M_n  -\frac{M_n^2}{2} + \ldots]
 \Big)_{II}
  } \cr
&& =
\delta_{II}  - g k_F [\frac{f_0}{2} \, n + f_1  (n-1) ] \delta z \, \delta_{II}     \cr
&&
=
e^{ - g k_F \big(\frac{f_0}{2}+ f_1 \big) |z| } \delta_{II}. 
\end{eqnarray}
In the third equality, we have dropped higher-order terms that vanish as $n \rightarrow \infty$.
Further details on such computations are given in Appendix \ref{Appendix-DisAvg}.
Defining 
\begin{align}
\label{lambdaandgeffdef}
g_{\rm eff} \equiv g \big({f_0 \over 2} + f_1\big),
\end{align}
we obtain the disorder-averaged Green's function,
\begin{eqnarray}
\label{Dis-Green-Realspace}
\overline{\langle  R_I(t,z) R^\dagger_I(0,0) \rangle}
=
\frac{i}{2\pi}  \frac{  e^{- g_{\rm eff} k_F  |z| }   }{(z - t) + i\alpha } .
\end{eqnarray}
This Green's function has a spatial decay length $\lambda = 1/g_{\rm eff} k_F$.

We now use the Green's function \eqref{Dis-Green-Realspace} to check that the density of states is finite.
For this, we need the retarded Green's function averaged over the disorder:
\begin{align}
\overline{G_{II}^R}(t,z;0,0) = - i \Theta(t) \overline{ \langle \{ R_I(t,z), R_I(0,0) \} \rangle } = - {i \over \pi} { \alpha e^{- g_{\rm eff} k_F  |z| } \over (z - t)^2 + \alpha^2},
\end{align}
where $\Theta(t)$ is the step function.
Fourier transforming $\overline{G^R}(t,z;0,0)$ for $\alpha \rightarrow 0$, we obtain
\begin{align}
\label{retardedfourier}
\overline{G_{II}^R}(\omega, p_\perp) = {1 \over \omega - p_\perp + i g_{\rm eff} k_F}.
\end{align}
From \eqref{retardedfourier}, we obtain the density of states per unit volume,
\begin{align}
\rho(\omega) = - {1 \over \pi} {\rm Im} \int {dp_\perp \over 2 \pi} {1 \over \omega - p_\perp + i g_{\rm eff} k_F} = {1 \over 2 \pi}.
\end{align}

\subsection{Longitudinal Conductivity}

We next turn to the disorder-averaged longitudinal conductivity $\sigma_{xx}(\omega)$.
The Kubo formula reads
\begin{align}
\label{kubo}
\sigma_{xx}(\omega) = {1 \over i \omega_n} {1 \over L} {k_F \over N} \int_0^\beta d \tau e^{i \omega_n \tau} \overline{\langle {\cal T}_\tau J_x(\tau) J_x(0) \rangle } \,\Big|_{\omega_n \rightarrow \omega + i 0^+},
\end{align}
where $\tau = i t$, $\omega_n = (2n+1)\pi \beta$ is a positive Matsubara frequency at temperature $1/\beta$, and $L \cdot N / k_F$ is a spatial volume factor equal to $\lim_{{\bf q} \rightarrow 0} \delta({\bf q}) = \lim_{p_\perp \rightarrow 0} \delta(p_\perp) \cdot {1 \over 2 \pi} \delta(k_F \theta_I)$ for some $I$.

Plugging in the expression for the current $J_x(\tau)$ \eqref{discretexcurrent} and using the unitary \eqref{rotationdiscrete}, the current two-point function is
\begin{align}
\label{current2point}
\langle {\cal T}_\tau J_x(\tau) J_x(0) \rangle & = \int dz dz' \sum_{I, J, A, B, C, D} \langle \tilde R^\dagger_A(\tau, z) \tilde R_B(\tau, z) \tilde R^\dagger_C(0, z') \tilde R_D(0, z') \rangle \cr
& \cdot \overline{U_{AI}(z) \cos(\theta_I) U^\dagger_{IB}(z) U_{CJ}(z') \cos(\theta_J) U^\dagger_{JD}(z') } \cr
& = \int dz dz' \, \Big({i \over 2 \beta \sinh\big( {\pi(z - z' + i (\tau - \tau') \over \beta } \big)} \Big)^2 \cdot {\rm Tr} \, \overline{ U(z) C U^\dagger(z) U(z') C U^\dagger(z') }  . \cr
\end{align}
In the first equality, we introduced the diagonal matrix $C_{I I'} = \cos\big({2 \pi I \over N} \big) \delta_{I I'}$; 
in the second equality, we used \eqref{discretetwo} to compute the finite-temperature fermion four-point function and expressed the product of $U$ and $C$ matrices using standard matrix notation. 
The disorder average of the product of $U$ and $C$ matrices is computed as in the previous section.
To this end, we partition the interval $[0, z]$ into $n$ segments and the interval $[0, z']$ into $m$ segments, each of size $\delta z = z/n = z'/m$, and decompose each $U$ matrix as in \eqref{Udecomposition}.
We find
\begin{align}
\label{Uproductaverage}
& {\rm Tr} \overline{ \big( e^{-i \, M_n }   \cdots e^{-i \, M_1  } \big)
 C
 \big( e^{ i \, M_1  } \cdots e^{ i \, M_n }   \big)
\big(  
 e^{-i \, M_m } \cdots  e^{-i \, M_1  }  \big)
\, 
C
\big(   e^{i \, M_1  } \cdots e^{ i \, M_m }  \big) } \cr
& = \sum_I \cos^2 \big({2 \pi I \over N} \big) e^{- 2 g_{\rm eff} k_F | z - z' |} \cr
& = {N \over 2} e^{- 2 g_{\rm eff} k_F | z - z' |},
\end{align}
where $g_{\rm eff}$ is defined in \eqref{lambdaandgeffdef}.
The details for the evaluation of the disorder average are given in Appendix \ref{Appendix-DisAvg}.
The sum over $I$ in the second equality is performed for $N \rightarrow \infty$ using the continuum limit ${1 \over N} \sum_I \rightarrow \int {d \theta \over 2 \pi}$.

Inserting \eqref{Uproductaverage} into the current two-point function \eqref{current2point}, the conductivity \eqref{kubo} becomes
\begin{align}
\sigma_{xx}(\omega) = {1 \over i \omega_n} {k_F \over 2 L} \int dz dz' d \tau \, e^{i \omega_n \tau}  \Big({i \over 2 \beta \sinh^2\big( {\pi(z - z' + i (\tau - \tau') \over \beta } \big)} \Big)^2 e^{- 2 g_{\rm eff} k_F | z - z' |} \Big|_{\omega_n \rightarrow \omega + i 0^+}.
\end{align}
Shifting $z \rightarrow z + z'$ and $\tau \rightarrow \tau + \tau'$, the integral over $z'$ produces a factor of $L$; we next calculate the integral over $\tau$ in the zero temperature limit $\beta \rightarrow \infty$.
The relevant term is
\begin{align}
\int_0^\infty d \tau e^{ i \omega_n \tau} \Big( {i \over 2 \pi} {1 \over \big( z + i \tau \big)} \Big)^2 = - {1 \over 4 \pi} | \omega_n|\, e^{- \omega_n z } \Theta\big(\omega_n z \big). 
\end{align}
Performing the remaining integral over $z$, we obtain the conductivity
\begin{align}
\label{conductivityresult}
\sigma_{xx}(\omega) = {1 \over 8} \cdot {1 \over g_{\rm eff} - i {\omega \over 2 v k_F}} \quad \text{(units of $e^2/h$)},
\end{align}
where we have restored the fermion velocity $v$, previously set to one.
This is the main result of this paper.
The dc longitudinal conductivity varies as $1/g_{\rm eff}$ along the random fixed line. 
An identical calculation with $\cos\big({2 \pi I \over N} \big) \rightarrow \sin\big({2 \pi I \over N} \big)$ produces $\sigma_{yy}(\omega) = \sigma_{xx}(\omega)$.
The Hall conductivity $\sigma_{xy}(\omega)$ vanishes because there is no time-reversal symmetry breaking on average.
This follows from an explicit computation similar to the above, in which $\sum_I \cos^2\big({2 \pi I \over N} \big) \rightarrow \sum_I \cos\big({2 \pi I \over N} \big) \sin\big({2 \pi I \over N} \big) = 0$.

We may crudely estimate the regime of validity of \eqref{conductivityresult} as follows.
Recall from \eqref{vectordisorderdef} that $g_{\rm eff} \sim g$ characterizes the scale of the fluctuations of the random vector potential ${\bf A}({\bf q})$, which in turn determines the random coupling $|V_{\theta \theta'}(p_\perp - p'_\perp)| \sim k_F |{\bf A}({\bf q})| \sim  k_F \sqrt{g_{\rm eff}}$ in \eqref{simplifiedlowES1}.
General effective field theory considerations require $k_F \sqrt{g_{\rm eff}} \leq \Lambda$, where $\Lambda \ll k_F$ is the cutoff on momenta transverse to the Fermi surface.
Inserting this inequality into \eqref{conductivityresult} at $\omega = 0$, we find
\begin{align}
\label{bound}
\sigma_{xx}(\omega = 0) \geq {1 \over 8} \cdot {k_F^2 \over \Lambda^2}.
\end{align}
$\Lambda \ll k_F$ (rather than $\Lambda \leq k_F$) ensures the scattering is primarily tangential to the Fermi surface, instead of perpendicular to it (see \S \ref{disordersection}).
A study of the effects of the various leading corrections to the effective theory \eqref{lowEbeforegauge} could potentially clarify the bound \eqref{bound}.

\section{Discussion and Summary} 
\label{discussion}

We have studied the effects of a quenched random, transverse magnetic field on a 2d Dirac fermion placed at finite density.
For weak disorder of sufficiently long wavelength, we showed how the effective theory reduces to an infinite collection of chiral fermions coupled by the vector potential.
This simplification allows for an exact treatment of the effects of the disorder.
We found a line of fixed points along which the longitudinal dc conductivity \eqref{conductivityresult} varies continuously with the disorder variance.

The dc conductivity was calculated in the collisionless $\hbar \omega/k_B T \rightarrow \infty$ limit.
It is important to extend our study to the opposite order of limits $\hbar \omega/k_B T \rightarrow 0$, the so-called incoherent regime  \cite{PhysRevB.57.7157}, relevant to experiment.
(See \cite{PhysRevB.75.205344} for a study of distinct dc limits of the ac conductivity of a clean Dirac fermion at zero density.)
This question is pertinent to the expected universality of the conductivity at a quantum phase transition \cite{SondhiGirvinCariniShahar}.
Numerical studies\footnote{We are grateful to Prashant Kumar for discussions about this.} \cite{PhysRevLett.70.481, KRAMER2005211, PhysRevLett.95.256805} of the integer quantum Hall transition appear to be roughly consistent with experiment (e.g., \cite{Shahar1995, Yang_2000}), giving a value for the dc longitudinal conductivity $\sigma_{xx} \sim (.54 - .60) \, e^2/h$.
%interactions important?

We focused exclusively on the point where the Dirac fermion is massless.
In the clean limit at finite density, a metal intervenes between integer quantum Hall states with $\sigma_{xy} = \pm {1 \over 2} {e^2 \over h}$ as the mass $m$ is tuned between $\pm \mu$, where $\mu$ is the chemical potential.
Based on numerics (e.g., \cite{PhysRevLett.126.076801}), the metallic region is absent in the presence of disorder and a direct integer quantum Hall transition should obtain.
It would be interesting to redo our analysis with a finite mass $m$ to try to find the localization length exponent for this transition.

The disorder we studied was of sufficiently long wavelength that it mediated elastic scattering between nearby Fermi points only.
The opposite regime, in which all Fermi points are coupled by the disorder, might be interesting to consider.
The action of the theory at energy $\omega = 0$ takes the form:
\begin{align}
\label{SYKlike}
S_{\omega = 0} = \int dz  \sum_{I, J =1}^N \, R^\ast_{I}(z) \Big(i \partial_z \delta_{IJ} + J_{I J} \Big) R_{J}(z),
\end{align}
with random $J_{IJ}$ coupling all Fermi points $I, J$, subject to a given ensemble.
Interpreting $z$ as ``time," this action is reminiscent of the quadratic Sachdev-Ye-Kitaev model \cite{kitaevbhmodel, PhysRevLett.70.3339} with complex fermions.

The similarity of the effective theory \eqref{discreteone} of the random Dirac Fermi surface to the theory of $N$ chiral free fermions in 1d suggests a possible route towards a non-Fermi liquid generalization, in which disorder may be studied simultaneously.
For example, we may consider two independent Dirac fermions---with chemical potentials that are of equal magnitude and opposite sign---in the presence of quenched vector potential disorder.
The effective action is $S_R + S_L$, where
\begin{align}
\label{rightfree}
S_{\rm R} & = \int dt dz \sum_{I, J =1}^N \, R^\ast_{I}(t, z) \Big(  i \big(\partial_t + \partial_z \big) \delta_{IJ} + {1 \over N} V_{I J}(z) \Big) R_{J}(t, z), \\
\label{leftfree}
S_{\rm L} & = \int dt dz \sum_{I, J =1}^N \,( L^\ast_{I}(t, z) \Big(  i \big(\partial_t - \partial_z \big) \delta_{IJ} - {1 \over N} V_{I J}(z) \Big) L_{J}(t, z), 
\end{align}
and $R_I$ ($L_J$) is the low-energy excitation about the Fermi surface defined by positive (negative) chemical potential.
Couplings between right ($R^\ast_I R_I$) and left ($L^\ast_J L_J$) densities leads to Luttinger liquid-like behavior.
So long as the couplings preserve a diagonal subgroup of the $U(N) \times U(N)$ symmetry of the Fermi surface, the interactions and disorder can be studied simultaneously.
The implications and possible microscopic origin of such a theory are unclear. 
 
\section*{Acknowledgments}

We thank Prashant Kumar and Sri Raghu for useful conversations.
C.-J. Lee was partially supported by the scholarship of the Taiwan Ministry of Education. 
This material is based upon work supported by the U.S. Department of Energy, Office of Science, Office of Basic Energy Sciences under Award No.~DE-SC0020007.

%\section*{Appendix}
%\addcontentsline{toc}{section}{Appendices}
%\renewcommand{\thesubsection}{\Alph{subsection}}
%\renewcommand{\theequation}{I\thechapter.\arabic{eqnarray}}
%\def\theequation{\thesubsection.\arabic{equation}}

\appendix
 
\section{Disorder Average of Products of $U$ and $C$ Matrices} 
\label{Appendix-DisAvg}

In this appendix, we detail the evaluation of the disorder average of products of $U$ and $C$ matrices, focusing on the product that appears in \eqref{current2point}:
\begin{align}
\label{thingtocompute}
{\rm Tr} \, \overline{ U(z) C U^\dagger(z) U(z') C U^\dagger(z') } \equiv \overline{U_{AI}(z) C_{I I'} U^\dagger_{I' B}(z) U_{B J}(z') C_{J J'} U^\dagger_{J' A}(z') },
\end{align}
where $U_{IJ}(z)$ is defined in \eqref{rotationdiscrete}, $C_{I I'} = \cos\big({2 \pi I \over N} \big) \delta_{I I'}$, and the sums over $A, I, I', B, J, J' = 1, \ldots, N$ are understood.
The computation of \eqref{averageoftwoUs} is similar and will not be discussed.

To calculate \eqref{thingtocompute}, we discretize the $z$ direction into segments $[z_{k-1}, z_{k}]$ of length $\delta z > 0$, where $z_k = k \delta z$ for all integer $k$. 
We take $z = z_n$, $z' = z_m$, and consider the limit $n, m \rightarrow \infty$ with $z_n$ and $z_m$ fixed.
$U(z)$ is decomposed as 
\begin{align}
\label{Udecompositionappendix}
U(z) = e^{- {i \over N} \int_{z_{n-1}}^{z_n} dz' \, V(z') } \cdots e^{- {i \over N} \int_{z_{1}}^{z_2} dz' \, V(z') } e^{- {i \over N} \int_{z_{0}}^{z_1} dz' \, V(z') }
\end{align}
and similarly for $U(z')$.
Since $\delta z$ is infinitesimal, we approximate 
\begin{align}
\label{integralapprox}
{1 \over N} \int_{z_{j-1}}^{z_j} dz' \, V(z') \approx {1 \over N} V(z_j) \delta z \equiv M_j.
\end{align}
The decomposition of $U$ becomes
\begin{align}
\label{UMdecomposition}
U(z) = e^{-i \, M_n } e^{- i M_{n-1}} \cdots e^{-i \, M_1}.
\end{align}
From \eqref{approxtVdisorderdefdiscrete}, $M_j$ is a zero-mean Gaussian random variable with variance,
\begin{align}
\overline{(M_i)_{IJ} (M_j)_{KL}} = {g k_F \over N} \delta_{IL} \delta_{JK} f(z_i - z_j) \delta z^2,
\end{align}
where $f(z_i - z_j)$ is defined by
\begin{eqnarray}
f(z_i-z_j) \delta z  =
 f_0 \delta_{|i-j|,0}
 +
 f_1 \delta_{|i-j|,1} 
 + f_2 \delta_{|i-j|,2}
 +... 
 + f_k \delta_{|i-j|,k}.
  \label{fzizj-choice}
\end{eqnarray}
We refer to $f_0$ as the on-site correlation coefficient, $f_1$ as the $1^{st}$ neighbor correlation coefficient,  $f_2$ as the $2^{nd}$ neighbor correlation coefficient, etc.
Using \eqref{fzizj-choice}, we write the disorder average \eqref{thingtocompute} as
\begin{align}
{\rm Tr} \, \overline{ U(z) C U^\dagger(z) U(z') C U^\dagger(z') } = \Tr[ C \, C ]   +     
W_0 + W_1+ \ldots + W_k,
\label{W0123-Decomp}
\end{align}
where $W_j$ denotes the contribution from $f_j$.  
The first term $ \Tr[ C \, C ] $ is the constant term without any Wick contraction due to the disorder averaging.

\subsection*{$1^{st}$-neighbor correlation: $W_1$}

We begin with the $1^{st}$ neighbor contribution $W_1$.
We substitute \eqref{UMdecomposition} into \eqref{W0123-Decomp} and expand the exponentials to obtain
\begin{align}
\label{W1}
W_1 & = \overline{ \Big( e^{-i \, M_n } \cdots e^{-i \, M_1  } \Big)_{AI}
 C_{II'}
 \Big( e^{ i \, M_1  } \cdots e^{ i \, M_n }   \Big)_{I'B} } \cr
 & \times \overline{ \Big( e^{-i \, M_m } \cdots e^{-i \, M_1  }  \Big)_{BJ}
\, 
C_{JJ'}
\Big(   e^{ i \, M_1  } \cdots e^{ i \, M_m }  \Big)_{J'A} } \Big{|}_{f_1} \cr
& \approx \overline{ \Big(  
( 1- iM_n ) \cdots (1- iM_{1} )
\Big)_{AI} C_{II'}
\Big(  
( 1+ iM_1 ) \cdots (1 +  iM_{n} )
\Big)_{I'B}  } \cr
& \times \overline{ \Big(  
( 1- iM_m ) \cdots (1- iM_{1} )
\Big)_{BJ} C_{JJ'}
\Big(  
( 1+ iM_1 ) \cdots (1 +  iM_{m} )
\Big)_{J'A} } \Big{|}_{f_1},
\end{align}
where $\Big{|}_{f_1}$ indicates that we only consider the contributions due to the $1^{st}$ neighbor correlation $f_1$ in  \eqref{fzizj-choice}.
Note that the disorder average in the first equality is being performed over the entire product, not separately over each term in the product.
Assuming $n > m$, there are four possible contractions involving $M_n$ that we isolate by writing $W_1$ as
\begin{align}
&  \overline{ \Big(   (-i M_{n}) (-i M_{n-1})  \Big)_{AI} }  
C_{II'} \delta_{I'B}  \; \delta_{BJ} C_{JJ'} \delta_{J'A} 
+ \overline{ (-i M_n)_{AI} (i M_{n-1})_{I' B}  }  
 \; C_{II'}  \delta_{BJ} C_{JJ'} \delta_{J'A}   \cr
& + \overline{ (-i M_{n-1})_{AI}(i M_n)_{I'B} } C_{II'}   
   \;  \delta_{BJ} C_{JJ'} \delta_{J'A}
+
\delta_{AI} C_{II'}   \overline{ 
      \Big((i M_{n-1}) (i M_{n}) \Big)_{I'B } } 
   \;  \delta_{BJ} C_{JJ'} \delta_{J'A}  
\cr
&  +
\overline{
\Big(  e^{-i \, M_{n-1}   } \cdots e^{-i \, M_1  } \Big)_{AI}
 C_{II'}
 \Big( e^{ i \, M_1  } \cdots e^{ i \, M_{n-1}   }    \Big)_{I'B}  
\Big(  
 e^{-i \, M_m } \cdots e^{-i \, M_1  }  \Big)_{BJ}
\, 
C_{JJ'}
\Big(   e^{ i \, M_1  } \cdots e^{ i \, M_m }  \Big)_{J'A} } \Big{|}_{f_1}   \cr
%%%%%%%%%%%%%%
&
=   
2  g\, f_1 \,k_F  \, \delta z\,
\Big(  
 -  \delta_{AI} \delta_{I'B}  + \frac{1}{N}   \delta_{II'} \delta_{AB}  
 \Big) \delta_{BJ}  \hat{C}_{II'} \hat{C}_{JJ'} \delta_{J'A}  \cr
 &
+
\overline{
\Big(  e^{-i \, M_{n-1}   } \cdots e^{-i \, M_1  } \Big)_{AI}
 C_{II'}
 \Big( e^{ i \, M_1  } \cdots e^{ i \, M_{n-1}   }    \Big)_{I'B}  
\Big(  
 e^{-i \, M_m } \cdots e^{-i \, M_1  }  \Big)_{BJ}
\, 
C_{JJ'}
\Big(   e^{ i \, M_1  } \cdots e^{ i \, M_m }  \Big)_{J'A} } \Big{|}_{f_1}. \cr
\end{align}
The subleading term $\propto \frac{1}{N}$, which comes from ``crossing-contractions," vanishes because $C_{I I'} \delta_{I I'} = \Tr[C] = 0$.

We continue as above sequentially contracting $1^{st}$ neighbor pairs involving $M_{n-1}, M_{n-2}, \ldots, M_{m + 2}$ to arrive at
\begin{align}
\label{W1uptoMm1}
&&
W_1  =
-2  g\, f_1 \,k_F  
 \Tr[ C C] \Big(  n- (m+2)+1   \Big) \; \delta z
+ \Big(  \text{Contractions of $M_{m+1}, \ldots ,M_{1}$} \Big) 
\end{align}
There are eight contractions that involve $M_{m+1}$ and $M_m$ in the second term in \eqref{W1uptoMm1}:
\begin{eqnarray}
&& \Big[
  \overline{ \Big(  (-i M_{m+1}) (-i M_{m}) \Big)_{AI} } C_{II'}  \delta_{I'B} 
\delta_{BJ} C_{JJ'} \delta_{J'A}
+
 \overline{ (-i M_{m+1})_{AI}  (i M_m)_{I'B} } C_{II'}\delta_{BJ} C_{JJ'} \delta_{J'A}  \cr
&&
+
\overline{ (-i M_{m+1})_{AI}  (-i M_m)_{BJ}  } C_{II'} \delta_{I'B}  C_{JJ'} 
\delta_{J'A} 
+
\overline{ (-i M_{m+1})_{AI}  (+i M_m)_{J'A} } C_{II'}  \delta_{I'B} 
\delta_{BJ} C_{JJ'}     \Big]   \cr
&&
+ \Big[
\overline{ (M_m)_{AI} (M_{m+1})_{I'B} } C_{II'} \delta_{BJ} C_{JJ'} \delta_{J'A}
- \overline{ (M_m M_{m+1})_{I'B} } \delta_{AI} C_{II'} 
  \delta_{BJ} C_{JJ'} \delta_{J'A}  \nn \\
&&
+ \delta_{AI} C_{II'} \overline{ (M_{m+1})_{I'B} (M_m)_{BJ} } C_{JJ'} \delta_{J'A}
-
\delta_{AI} C_{II'} \overline{ (M_{m+1})_{I'B} (M_m)_{J'A} } \delta_{BJ}  C_{JJ'} 
\Big] .
\end{eqnarray}
In the first $[\ldots]$, the first term cancels with the fourth and the second term cancels with the third; the same pairs mutually cancel in the second $[\ldots]$.
Next we consider the contractions involving $M_m$ and $M_{m-1}$, and we again obtain zero.
These cancellations continue through to contractions involving $M_1$.
This makes sense since we expect the result to only depend on the difference $(n-m)$.
Summarizing, we have
\begin{eqnarray}
W_1 
= -
 \Tr[C C]
\Big( 2  g\, f_1 \,k_F    \,  [  n- (m+2)+1   ] \delta z \Big).
\end{eqnarray}

\subsection*{$2^{nd}$-neighbor correlation and beyond: $W_2, \ldots, W_k$}

Following the logic used to compute the $1^{st}$ neighbor correlations, we find
\begin{align}
W_2 & = \overline{ \Big( e^{-i \, M_n } \cdots e^{-i \, M_1  } \Big)_{AI}
 C_{II'}
 \Big( e^{ i \, M_1  } \cdots e^{ i \, M_n }   \Big)_{I'B} } \cr
 & \times \overline{ \Big( e^{-i \, M_m } \cdots e^{-i \, M_1  }  \Big)_{BJ}
\, 
C_{JJ'}
\Big(   e^{ i \, M_1  } \cdots e^{ i \, M_m }  \Big)_{J'A} } \Big{|}_{f_1} \cr
& \approx \overline{ \Big(  
( 1- iM_n ) \cdots (1- iM_{1} )
\Big)_{AI} C_{II'}
\Big(  
( 1+ iM_1 ) \cdots (1 +  iM_{n} )
\Big)_{I'B}  } \cr
& \times \overline{ \Big(  
( 1- iM_m ) \cdots (1- iM_{1} )
\Big)_{BJ} C_{JJ'}
\Big(  
( 1+ iM_1 ) \cdots (1 +  iM_{m} )
\Big)_{J'A} } \Big{|}_{f_2} \cr
& = - \Tr[ C C]
\Big(    2  g\, f_2 \,k_F    \,  [  n- (m+3)+1   ] \delta z \Big) .
\end{align}
Generalizing to $k^{th}$ neighbor correlations, we have
\begin{align}
W_k
= -  \Tr[C C]
\Big(    2  g\, f_k \,k_F    \,  [  n- (m+k+1)+1   ] \delta z \Big) .
\end{align}

\subsection*{On-site Correlation: $W_0$}
Unlike the off-site correlations discussed above, we have to expand the exponentials to quadratic order to obtain the contribution from on-site correlations:
\begin{align}
W_0 & = \overline{ \Big( e^{-i \, M_n } \cdots e^{-i \, M_1  } \Big)_{AI}
 C_{II'}
 \Big( e^{ i \, M_1  } \cdots e^{ i \, M_n }   \Big)_{I'B} } \cr
 & \times \overline{ \Big( e^{-i \, M_m } \cdots e^{-i \, M_1  }  \Big)_{BJ}
\, 
C_{JJ'}
\Big(   e^{ i \, M_1  } \cdots e^{ i \, M_m }  \Big)_{J'A} } \Big{|}_{f_0} \cr
& \approx
  \overline{
\Big(  
( 1- iM_n -\frac{M_n^2}{2}) \cdots (1- iM_{1} -\frac{M_1^2}{2})
\Big)_{AI}  \hat{C}_{II'}
\Big(  
( 1+ iM_1 -\frac{M_1^2}{2}) \cdots (1+ iM_{n} -\frac{M_n^2}{2} )
\Big)_{I'B} } \cr
& \times
\overline{ \Big(  
( 1+ iM_m -\frac{M_m^2}{2} ) \cdots (1+  iM_{1} - \frac{M_1^2}{2}  )
\Big)_{BJ}  \hat{C}_{JJ'}
\Big(  
( 1+ iM_1 -\frac{M_1^2}{2}) \cdots (1 +  iM_{m} -\frac{M_m^2}{2} )
\Big)_{J'A} }  \Big{|}_{f_0} \cr
& = g f_0 k_F \delta z \Big(  
 -  \delta_{AI} \delta_{I'B}  + \frac{2}{N}   \delta_{II'} \delta_{AB}  
 \Big) \delta_{BJ}  \hat{C}_{II'} \hat{C}_{JJ'} \delta_{J'A} 
 +
 \Big(  \text{Contractions of $M_{n-1}, \ldots, M_1$} \Big) . \cr
\end{align}
The crossing term $\propto {1 \over N}$ again vanishes.
Continuing in this way we arrive at 
\begin{align}
W_0 
=
- \Tr[ C C]
\Big(  g\, f_0 \,k_F    \,  [  n- (m+1)+1   ] \delta z \Big)
+
 \Big(  \text{Contractions of $M_{m}, \ldots, M_1$} \Big) .
\end{align}
Similar to the cancellations that were discussed in the context of the $1^{st}$ neighbor correlations, the remaining contractions of $M_m, \ldots, M_1$ equal zero.
Thus, we have
\begin{align}
W_0 
=
- \Tr[ C C]
\Big(  g\, f_0 \,k_F    \,  [  n- (m+1)+1   ] \delta z \Big) .
\end{align}

\subsection*{Final Result}

Collecting all correlations $f_0, f_1, \ldots, f_k$ and replacing $\lim_{n \rightarrow \infty} n \delta z = z$ and $\lim_{m \rightarrow \infty} m \delta z = z'$, we have
\begin{align}
{\rm Tr} \, \overline{ U(z) C U^\dagger(z) U(z') C U^\dagger(z') } & = \Tr[C C] \Big( 1 - 2 g k_F \delta z \big({f_0 \over 2} (n - m) +  \sum_{j = 1}^k f_j [n - m - k ] \big) \Big) \cr
& = \Tr[C C] e^{- 2 g k_F |z - z'| \big({f_0 \over 2} + f_1 + \ldots f_k \big)}.
\end{align}
Note that each term proportional to $k$ in the sum vanishes at large $n$ and $m$.
This concludes our calculation of the disorder average in \eqref{Uproductaverage}.
In the main text, $f_0$ and $f_1$ are nonzero, and $f_{k \geq 2} = 0$.

\bibliographystyle{utphys}  % TO use this, we have to put the file utphys.bst in the same director of this tex file

%\bibliography{myBibDataBase}{}   
\bibliography{myBibDataBase}{} % file name of   myBibDataBase.bib
% Compile steps :  pdflatex -> bibtex -> pdflatex -> pdflatex

%\cite{PhysRevResearch.2.023303}   % Make a test 
%\nocite{*} % to test all bib entrys

% Run the compiling of "BibTex" in the main latex file (i.e. this file), and then run
% "fast compiling" for regular compiling step I do usually.   
% There is no need to do any compilation for by data base XXX.bib  file

\end{document}